\documentclass[12pt]{article}
\usepackage{graphicx}
\usepackage{nips14submit_e,times,amsmath,amssymb,graphicx}
\usepackage{algpseudocode}
\usepackage{algorithm}
\usepackage{subfigure}
\usepackage{comment}

\RequirePackage[numbers]{natbib}
\RequirePackage[colorlinks,citecolor=blue,urlcolor=blue]{hyperref}

\title{Generalized Mode and Ridge Estimation}

\author{
Yen-Chi Chen\\
Department of Statistics\\
Carnegie Mellon University\\
\texttt{yenchic@andrew.cmu.edu} \\
\And
Christopher R. Genovese\\
Department of Statistics\\
Carnegie Mellon University\\
\texttt{genovese@stat.cmu.edu}\\
\And
Larry Wasserman\\
Department of Statistics\\
Carnegie Mellon University\\
\texttt{larry@stat.cmu.edu}\\
}

\newcommand\given{\mid}

\newcommand\cL{\mathcal{L}}
\newcommand\cM{\mathcal{M}}
\newcommand\cR{\mathcal{R}}

\nipsfinalcopy

\begin{document}
\maketitle{}

\begin{abstract}
The generalized density is a product of
a density function and a weight function.
For example,
the average local brightness of an astronomical image
is the probability of finding a galaxy times the mean brightness of the galaxy.
We propose a method for studying the geometric structure of
generalized densities.
In particular,
we show how to find the modes and ridges of a generalized density function
using a modification of the mean shift algorithm and its variant, subspace constrained
mean shift.
Our method can be used to perform clustering
and to calculate a measure of connectivity between clusters.
We establish consistency and rates of convergence for our estimator
and apply the methods to data from two astronomical problems.
\end{abstract}

\newtheorem{thm}{Theorem}
\newtheorem{lem}[thm]{Lemma}
\newtheorem{cor}[thm]{Corollary}
\newtheorem{proposition}[thm]{Proposition}
\newenvironment{definition}[1][Definition]{\begin{trivlist}

\item[\hskip \labelsep {\bfseries #1}]}{\end{trivlist}}
\let\hat\widehat
\setlength{\parindent}{0cm}
\setlength{\parskip}{\baselineskip}

\newcommand\R{\mathbb{R}}
\newcommand\E{\mathbb{E}}
\newcommand\mathand{\ {\rm and}\ }
\newcommand\norm[1]{\|#1\|}
\newcommand\dest{{\sf dest}}
\newcommand\MISE{{\sf MISE}}
\newcommand\mode{{\sf Mode}}
\newcommand\ridge{{\sf Ridge}}
\newcommand\Cov{{\sf Cov}}
\newcommand\Var{{\sf Var}}

\catcode`@=11
\newskip\beforeproofvskip
\newskip\afterproofvskip
\beforeproofvskip=\medskipamount
\afterproofvskip=\medskipamount
\def\proofsquare{\square}
\def\prooftag{Proof}
\def\proofskip{\enspace}

\def\proof{\@ifnextchar[{\@@proof}{\@proof}}  
\def\@startproof{\par\vskip\beforeproofvskip\leavevmode}
\def\@proof{\@startproof{\scshape\prooftag.}\proofskip}
\def\@@proof[#1]{\@startproof {\scshape\prooftag #1.}\proofskip}
\def\endproof{\hskip 1em $\proofsquare$\par\vskip\afterproofvskip}
\catcode`@=12

\section{Introduction}

Consider a random sample of the form $(X_1,Y_1), \ldots (X_n,Y_n)$,
where $X_1,\cdots,X_n\in\R^d$ is a random sample from a smooth
density $p$ and each $Y_i$ is a scalar random variable.  The
\emph{generalized density function (GDF)},
also known as an intensity function,
is
$f(x) = \mu(x) p(x)$
where
$\mu(x)=\E\left(Y \given X=x\right)$.
A kernel estimate of the GDF is
\begin{equation}
\hat{f}_n(x) = \frac{1}{nh^d}\sum_{i=1}^n Y_iK\left(\frac{x-X_i}{h}\right),
\label{eq::I1}
\end{equation}
where $K$ is a kernel function and $h$ is the smoothing bandwidth.
For simplicity, we henceforth assume that the function $K$ is a
Gaussian kernel.  
The generalized density function and estimator are of interest in problems
where the additional information in the $Y$'s measure the importance, relevance,
or intensity of the corresponding points.
(In the language of point processes, the $X$'s are the points and the $Y$'s
are the marks.)
Two common cases include:
\begin{itemize}
\item[1.] $Y_i$ represents a covariate associated with each point, and
the generalized density weights the points according to the value of the covariate.
For example, in galactic astronomy, the $X$'s might represent a galaxy's location
and the $Y$'s the galaxy's mass. Astronomers are interested in the ``mass-density'' $f$,
which describes the distribution of galaxy mass.

\item[2.] $Y_i$ represents the measurement precision for each observation,
so the generalized density weights the points according to how precisely they are measured.
\end{itemize}

This paper focuses on estimating the \emph{modes} and \emph{ridges} of
the GDF because (i)~they are often features
of direct scientific interest, (ii)~they provide useful and
descriptive summaries of the GDF' structure,
and (iii)~they can be used as inputs to clustering.

Given a smooth function $f:\R^d\mapsto \R$, the modes (local maximums)
and ridges \cite{Eberly1996, Ozertem2011, Genovese2012a} are
defined as
\begin{equation}
	\begin{aligned}
	\cM=\mode(f) &= \{x\in\R^d: \nabla f(x)=0, \lambda_1(x)<0\}\\
	\cR=\ridge(f) & = \{x\in\R^d: V(x)V(x)^T \nabla f(x) = 0, \lambda_2(x)<0\},
	\end{aligned}
\end{equation}
where $\lambda_1(x)\geq \lambda_2(x)\geq\cdots\geq\lambda_d(x)$ is the
eigenvalues of $\nabla \nabla f(x)$, the Hessian matrix of $f$, and
$V(x) = [v_2,\cdots,v_d(x)]$ is a $d\times (d-1)$ matrix with $v_k(x)$
being eigenvectors of $\nabla\nabla f(x)$ corresponding to eigenvalue
$\lambda_k(x)$.

The problem of estimating modes and ridges for density functions has been considered
in the literature.ma
For instance, \cite{Romano1988, Mammen1992, Abraham2003} develop estimators for local modes,
and \cite{Genovese2012a} develop estimators for density ridges and establish the 
asymptotic properties of these estimators.
The analogous problem for generalized density functions has not yet been considered.

\begin{figure}
\centering
	\subfigure[Modes and ridges]
	{
		\includegraphics[width=1.5 in, height=1.5 in]{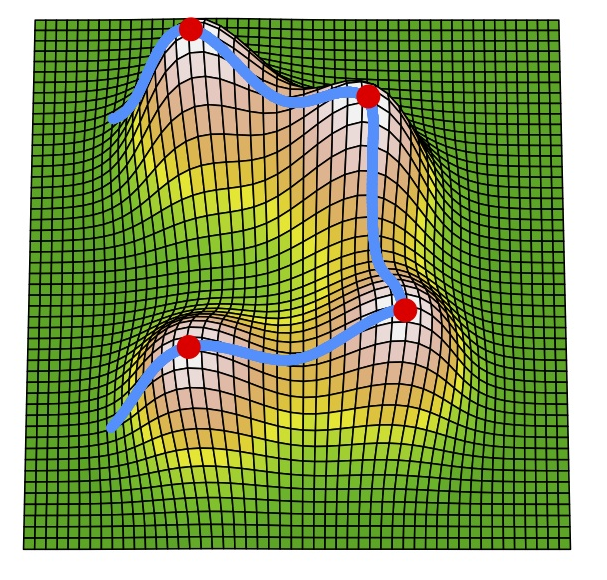}
	}
	\subfigure[Contour plot]
	{
		\includegraphics[width=1.5 in, height=1.5 in]{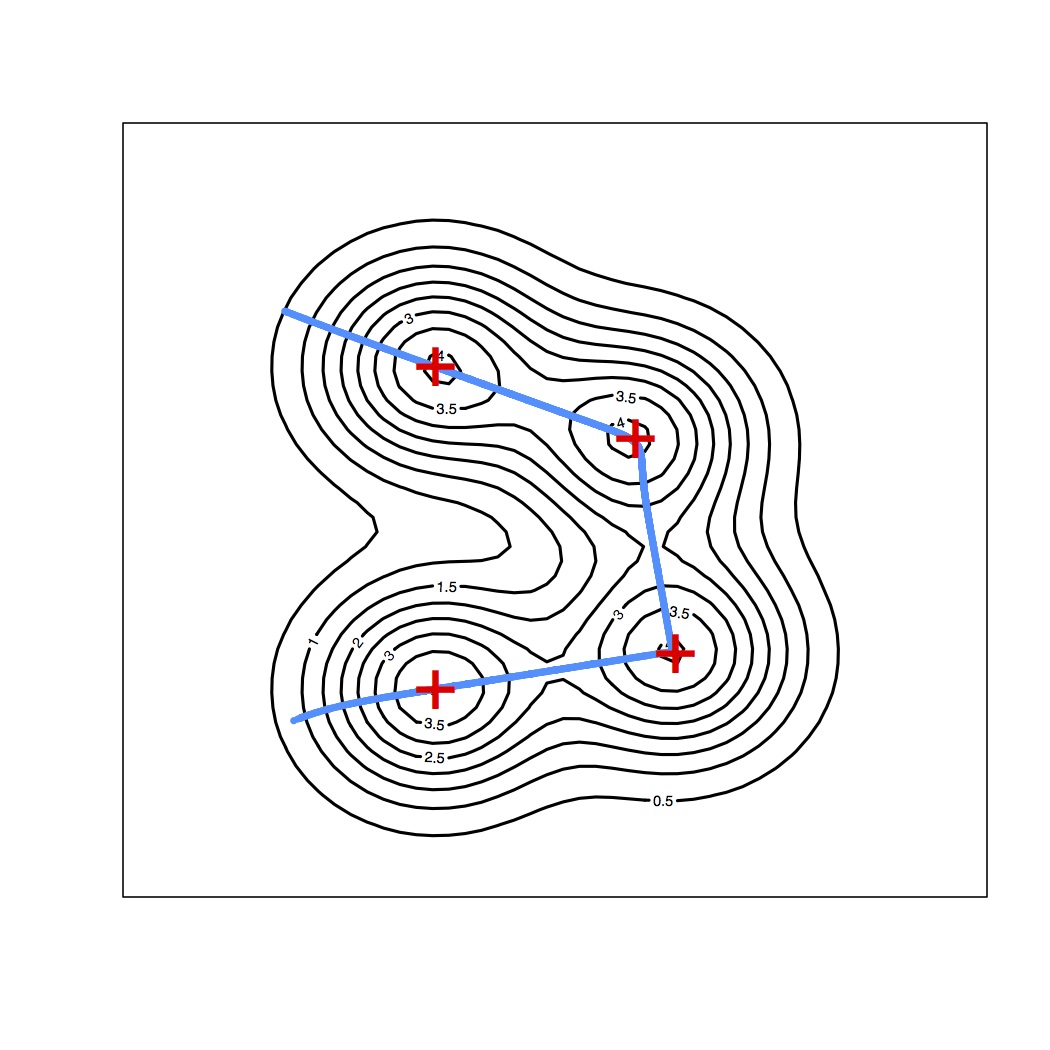}
	}
\caption{An example for a smooth function with its modes (red) and ridge (blue). }
\label{fig::ex0}
\end{figure}

We estimate the modes and ridges by
\begin{equation}
	\hat{\cM}_n=\mode(\hat{f}_n),\quad
	\hat{\cR}_n=\ridge(\hat{f}_n).
\end{equation}
It is well-known the local modes and ridges from the kernel density
estimator can be estimated efficiently by the mean shift
algorithm~\cite{Fukunaga,cheng1995mean,Comaniciu2002}.  Here, we
present a modification of the mean shift algorithm that can find the
modes and ridges in $\hat{f}_n$.

\section{Methods}

\subsection{Weighted Mean Shift}

Before we proceed to our method, we first review the usual mean shift algorithm.
Given data $X_1,\cdots,X_n\in\mathbb{R}^d$ and an initial point $x\in\mathbb{R}^d$,
the mean shift algorithm~\cite{Fukunaga,cheng1995mean,Comaniciu2002} 
updates $x$ to
\begin{equation}
x\longleftarrow \frac{\sum_{i=1}^n X_i 
K\left(\frac{x-X_i}{h}\right)}{\sum_{i=1}^n K\left(\frac{x-X_i}{h}\right)}.
\end{equation}
If we keep iterating, we end up at a mode of
the kernel density estimator 
$\hat{p}_n(x) = \frac{1}{nh^d}\sum_{i=1}^n K\left(\frac{x-X_i}{h}\right)$.

We define the
\emph{weighted mean shift}, which
generates a path starting from a point $x$ by successive updates
of the form
\begin{equation}
\boxed{
x\longleftarrow \frac{\sum_{i=1}^n Y_iX_i 
K\left(\frac{x-X_i}{h}\right)}{\sum_{i=1}^n Y_iK\left(\frac{x-X_i}{h}\right)}.}
\label{eq::wMS1}
\end{equation}
This is directly analogous to the ordinary mean-shift update but puts
additional weight $Y_i$ on the point $X_i$.  The path generated from
each point $x$ eventually converges to a local mode of the generalized
density estimate $\hat{f}_n$, an element of the set $\hat{\cM}_n$.
Later we will see that this is a consistent estimator of $\cM$.

Now we derive the relation between \eqref{eq::I1} and \eqref{eq::wMS1}. 
The gradient of $\hat{f}_n(x)$ is 
\begin{equation}
	\begin{aligned}
	\nabla \hat{f}_n(x) & = \frac{1}{nh^{d}}\sum_{i=1}^n Y_i \nabla K\left(\frac{x-X_i}{h}\right)\\
	&=\frac{1}{nh^{d+2}}\sum_{i=1}^n Y_i (X_i-x) K\left(\frac{x-X_i}{h}\right)\\
	& = \frac{1}{nh^{d+2}} \left(\sum_{i=1}^n Y_i X_i K\left(\frac{x-X_i}{h}\right)\right)-\frac{x}{nh^{d+2}}\left(\sum_{i=1}^n Y_i  K\left(\frac{x-X_i}{h}\right)\right).
	\end{aligned}
\end{equation}
Note that we use the fact that $\nabla K(x) = -x K(x)$ for the Gaussian kernel. After rearrangement,
\begin{equation}
x+m(x) = \frac{\sum_{i=1}^n Y_iX_i K\left(\frac{x-X_i}{h}\right)}{\sum_{i=1}^n Y_iK\left(\frac{x-X_i}{h}\right)},
\end{equation}
where 
\begin{equation}
m(x) = \frac{nh^{d+2}}{\sum_{i=1}^n Y_iK\left(\frac{x-X_i}{h}\right)} \times \nabla \hat{f}_n(x)
\label{eq::wMS4}
\end{equation}
is the mean shift vector that is always pointing toward the direction of gradient.
Thus, the update rule \eqref{eq::wMS1} is to move $x$ to $x+m(x)$ which 
follows the gradient ascent. 
By Morse theory, for a smooth function with non-degenerate Hessian matrix,
the gradient ascent path will converge to one of the local modes.
Accordingly, the update rule \eqref{eq::wMS1} ends up at one of the local modes.

\subsection{Weight Subspace Constrained Mean Shift}

\cite{Ozertem2011} proposes the {subspace-constrained mean-shift}
algorithm which can be used to find the ridges of $\hat{f}_n(x)$.  An
analogous modification also works in the weighted case to find the
ridges of the generalized density function.  This update rule, which
we call the \emph{weighted subspace constrained mean shift} algorithm,
is given by
\begin{equation}
\boxed{ x\longleftarrow x+V(x)V(x)^T m(x),}
\end{equation}
where $m(x)$ is the mean shift vector defined in \eqref{eq::wMS4} and
$V(x)= [v_2(x),\cdots,v_d(x)]$ with $v_k(x)$ being the eigenvector
corresponding to the $k$-th eigenvalue (first is the largest) of the
estimated Hessian matrix $\nabla \nabla \hat{f}_n(x)$. The Hessian
matrix is
\begin{equation}
\nabla \nabla \hat{f}_n(x) = 
\frac{1}{nh^{d+4}}\sum_{i=1}^n  \left((X_i-x)(X_i-x)^T -h^2\mathbf{I}_d\right) Y_iK\left(\frac{x-X_i}{h}\right),
\end{equation}
where $\mathbf{I}_d$ is the $d\times d$ identity matrix. In practice,
we can ignore the factor $ \frac{1}{nh^{d+4}}$ since it is just a
scaling.
This algorithm will push every point $x$ along a `projected gradient
path' $V(x)V(x)^T m(x)$ until it arrives a point on $\hat{\cR}_n$.

\section{Applications}

\subsection{Mode Clustering}

A common application of the mean shift algorithm is to perform a type
of clustering called \emph{mode clustering} \cite{li2007nonparametric}.  
The clusters are defined as the sets of points whose
mean-shift paths converge to the same local mode.  Using the weighted
mean-shift algorithm yields a mode clustering based on the generalized
density estimate.

Let $f(x)$ be a smooth intensity function. For any point $x$, we
define a gradient path $\phi_x(t)$ as follows
\begin{equation}
\phi_x: \R^{+}\mapsto \R^d,\quad
\phi_x(0)=x,\quad
\phi_x'(t) = \nabla f(\phi_x(t)).
\end{equation}
We denote $\dest(x) = \lim_{t\rightarrow \infty} \phi_x(t)$ as the
destination of $\phi_x$. By Morse theory \cite{Guest2001,Sousbie2011}, 
$\dest(x)$ must be one of the local modes of $f(x)$
except the case $x$ is in a set of Lebesque measure $0$
(including saddle points and local minimums). 
Let
$\cM=\{M_1,\cdots,M_k\}$ be the collection of local modes of $f$. We
define the cluster of $M_j$ by
\begin{equation}
C_j = \{x\in\R^d: \dest(x) =M_j\}.
\end{equation}

In practice, we cluster data points by their destination of the
weighted mean shift \eqref{eq::wMS1}.  

\subsection{Connectivity Measure for Clusters}

Some clusters are fairly isolated while other are close together.
Here we show how to measure how close clusters are by
defining a notion of connectivity.
The idea is that the mean shift iterations can be thought of
particles moving according to a
diffusion (Markov chain).

We define a diffusion as follows.
The probability of jumping from $x$ to $X_i$ is
\begin{equation}
\mathbb{P}(x\rightarrow X_i) = 
\frac{Y_iK\left(\frac{x-X_i}{h}\right)}{\sum_{i=1}^n Y_iK\left(\frac{x-X_i}{h}\right)}.
\label{eq::MS2}
\end{equation}
This defines a diffusion and we denote $Q(x)$ be the random variable of the above diffusion. i .e.
$\mathbb{P}(Q(x) = X_i) = \mathbb{P}(x\rightarrow X_i).$
Now
\begin{equation}
\E\left(Q(x)|X_1,Y_1,\cdots,X_n,Y_n\right) = \frac{\sum_{i=1}^n X_iY_i
  K\left(\frac{x-X_i}{h}\right)}{\sum_{i=1}^n
  Y_iK\left(\frac{x-X_i}{h}\right)}
\end{equation}
which is the update rule \eqref{eq::wMS1}.  The same result holds for
non-weighted mean shift.  Thus, the mean shift can be viewed as a
expectation for a certain diffusion.

Let $\hat{M}_1,\cdots,\hat{M}_k$ be the local modes of $\hat{f}_n(x)$.
Motivated by
\eqref{eq::MS2}, we define 
\begin{equation}
	\begin{aligned}
	\mathbb{P}(X_i\rightarrow X_j) &=
        \frac{Y_jK\left(\frac{x-X_j}{h}\right)}{\sum_{i=1}^n
          Y_iK\left(\frac{x-X_i}{h}
          \right)+\sum_{j=1}^kW_jK\left(\frac{x-\hat{M}_j}{h}\right)}\\ \mathbb{P}(X_i\rightarrow
        \hat{M}_j) &=
        \frac{W_jK\left(\frac{x-M_j}{h}\right)}{\sum_{i=1}^n
          Y_iK\left(\frac{x-\hat{M}_j}{h}
          \right)+\sum_{j=1}^kW_jK\left(\frac{x-\hat{M}_j}{h}\right)},
	\end{aligned}
\end{equation}
where $W_j = \hat{f}_n(\hat{M}_j)/\left(\frac{1}{nh^d}\sum_{i=1}^n
K\left(\frac{\hat{M}_j-X_i}{h}\right)\right)$. Note that we impute the
weight $W_j$ for each local mode $\hat{M}_j$ by an estimate
of $m(\hat{M}_j)$.  We also define each mode to be an absorbing
state. Namely, the transition probability to/from each local mode to
itself is $1$.

Let $\mathbf{P}$ be a transition matrix with $k+n$ states such that
the first $k$ states are the estimated local modes and the latter $n$
states are the data points. Then the transition matrix $\mathbf{P}$
can be factorized by
\begin{equation}
\mathbf{P} =\begin{bmatrix}
	\mathbf{I}_k& 0\\
	S&T
	\end{bmatrix},
\quad T_{ij} = \mathbb{P}(X_i\rightarrow X_j), 
\quad S_{ij} = \mathbb{P}(X_i\rightarrow \hat{M}_j),
\end{equation}
and $\mathbf{I}_k$ is the $k\times k$ identity matrix.

Then the matrix $A = \left(\mathbf{I}_n-T\right)^{-1}S$ is the
absorbing matrix; that is, the absorbing probability from $X_i$ to the
local mode $M_j$ is $A_{ij}$. We define the \emph{connectivity} for
the two clusters corresponding to local modes $M_i, M_j$ as
\begin{equation}
\Omega_{ij} = \frac{1}{2}\left(\frac{\sum_{X_l \in D_i}Y_l
  A_{lj}}{\sum_{X_l \in D_i}Y_l}+\frac{\sum_{X_l \in D_j}Y_l
  A_{li}}{\sum_{X_l \in D_j}Y_l}\right),
\end{equation}
where $D_l$ is the data points belonging to cluster $l$.

The interpretation of $\Omega_{ij}$ is as follows.
$\Omega_{ij}$ is the average hitting probability from
points in cluster $i$ that end up at mode $j$ first, and vice versa.
Connectivity will be large when two clusters are close and 
the boundary between them has high density. 
If we think of the (hard) cluster assignments as class labels,
the connectivity is analogous to the mis-classification rate between class $i$ and class $j$.

\section{Statistical Analysis}

The modes and ridges of $\hat f_n$
are estimates of the modes and ridges of $f$.
In this section we study the statistical properties of these estimators.
Let $\mathbf{BC}^k$ be the
set of bounded, $k$ times continuously differentiable
functions.
Let $\norm{A}_{\max}$ be the max norm of a vector or a matrix $A$. For
a smooth function $f$, we define the following operators.
A vector $\alpha =
(\alpha_1,\ldots,\alpha_d)$ of non-negative integers is called a
multi-index with $|\alpha| = \alpha_1 + \alpha_2 + \cdots + \alpha_d$
and corresponding derivative operator
$$
D^\alpha = \frac{\partial^{\alpha_1}}{\partial x_1^{\alpha_1}} \cdots 
\frac{\partial^{\alpha_d}}{\partial x_d^{\alpha_d}},
$$
where $D^\alpha f$ is often written as $f^{(\alpha)}$.
Namely, each $f^{(\alpha)}(x)$ is a $|\alpha|$-th order partial derivative of $f$.
For $j = 0,\ldots, 3$, define the following norms associated with derivatives
\begin{equation}
\norm{f}_{\infty}^{(j)} = \underset{\alpha:\; |\alpha| = j}{\max} \underset{x\in\R^d}{\sup} |f^{(\alpha)}(x)|,
\qquad \norm{f}^*_{\infty, k} = \underset{j=0,\cdots,k}{\max} \norm{f}^{(j)}_{\infty}.
\end{equation}
Note that for two functions $f_1,f_2$, $\norm{f_1-f_2}^*_{\infty, k}$
is a norm that measures
the differences between $f_1,f_2$ to the $k$-th order differentiation.

\textbf{General Assumptions.}
\begin{itemize}
\item[(A1)] The random variable $Y$ is uniformly bounded by a constant $C_Y<\infty$.
\item[(A2)] The function $f\in\mathbf{BC}^3$ and is a Morse function 
(the Hessian matrix is non-singular at all critical values).
\item[(K1)] The kernel function $K\in\mathbf{BC}^3$ and is symmetric, non-negative and 
$$\int x^2K^{(\alpha)}(x)dx<\infty,\qquad \int \left(K^{(\alpha)}(x)\right)^2dx<\infty
$$ 
for all $|\alpha|=0,1,2,3$.
\item[(K2)] The kernel function satisfies condition $K_1$ of
  \cite{Gine2002}. That is, there exists some $A,v>0$ such that for
  all $0<\epsilon<1$,
$\sup_Q N(\mathcal{K}, L_2(Q), C_K\epsilon)\leq \left(\frac{A}{\epsilon}\right)^v,$
where $N(T,d,\epsilon)$ is the $\epsilon-$covering number for a
semi-metric space $(T,d)$ and
$$
\mathcal{K} = 
\Biggl\{u\mapsto K^{(\alpha)}\left(\frac{x-u}{h}\right)
: x\in\R^d, h>0,|\alpha|=0,1,2,3\Biggr\}.
$$
\end{itemize}
Assumptions (A1-2) are mild regularity conditions.
Condition (K1) is common for a kernel function and (K2) is the smoothness condition
for the kernel function. In particular, the Gaussian kernel and any smooth kernel with compact
support satisfies (K1-2).

\subsection{Risk of $\hat f_n$}

Define
the mean integrated square errors (MISE) by
\begin{equation}
	\MISE_k(\hat{f}_n) = \E\int \sum_{|\alpha|=k}\left| \hat{f}^{(\alpha)}_n(x)- f^{(\alpha)}(x)\right|^2dx,
\end{equation}
for $k= 0,1,2,3$.
Note that as $k=0$ we obtain the usual MISE for $\hat{f}_n$. 
This is just an extension to higher order derivatives.

\begin{thm}
\label{thm::MISE}
Assume (A1-2) and (K1). Then
\begin{align*}
\MISE_k(\hat{f}_n) &= O\left(h^4\right)+O\left(\frac{1}{nh^{d+2k}}\right), \quad k=0,1,2,3.
\end{align*}
\end{thm}
\cite{Chacon2011} proves the above
Theorem for usual kernel density estimation.
We omit the proof
as it is similar to their proof.

We also have the following uniform bound. 
\begin{thm}
\label{thm::unif}
For a smooth function $f$, 
let $\norm{f}^*_{\infty,k}$ be defined as the above.
Assume (A1-2) and (K1-2). Then
\begin{align*}
\norm{\hat{f}_n-f}^*_{\infty,k} = O\left(h^2\right)+O_P\left(\sqrt{\frac{\log n}{nh^{d+{2k}}}}\right), 
\quad k=0,1,2,3.
\end{align*}
\end{thm}
The proof is essentially the same as \cite{Gine2002, Einmahl2005} by noting that
the random variable $Y$ is bounded by $C_Y$.
Similar results in kernel density estimation can be seen in \cite{Gine2002, Einmahl2005}.

\subsection{Mode Estimation and Ridge Recovery}

In this section we assume that the density is supported on a compact
subset $\mathbb{K}\subset\R^d$.  
For two sets $A,B$, the
{\em Hausdorff distance} is given by
\begin{equation}
d_H(A,B) = \inf\{r: A\subset B\oplus r, B\subset A\oplus r\},
\end{equation}
where $A\oplus r = \{x: d(x,A)\leq r\}$ 
and 
$d(x,A) = \inf_{y\in A}||x-y||_2$.

\begin{thm}
\label{thm::modes}
Let $\cM,\hat{\cM}_n$ be the collection of local modes of $f$ and
$\hat{f}_n$ respectively. Assume (A1-2), (K1-2) and
\begin{itemize}
\item[(M)] there exists $\lambda_0,\lambda_1,\delta>0$ such that
$\{x: \lambda_1(x)<-\lambda_1, \norm{\nabla f(x)}_{2}<\lambda_0\}\subset \cM\oplus \delta,$
where $\lambda_1(x)$ is the first eigenvalue to the Hessian matrix of $f(x)$.
\end{itemize}
When $\norm{\hat{f}_n-f}^*_{\infty,2}$ is sufficiently small,
\begin{align*}
d_H\left(\hat{\cM}_n, \cM\right) = O\left(h^2\right)+O_P\left(\sqrt{\frac{1}{nh^{d+2}}}\right).
\end{align*}
\end{thm}
The proof of Theorem~\ref{thm::modes} is in the supplementary
material. Here we present an outline for the proof.

{\sc Proof Outline.}
Let $\hat{M}_1,\cdots,\hat{M}_{\hat{k}}$ be the estimated local modes. 
Note that the number of estimated modes might be different from the number of true modes.
However, by assumption (M) and the fact $\norm{\hat{f}_n-f}^*_{\infty,2}$ is sufficiently small,
we have $\hat{\cM}_n\subset \cM\oplus\delta$ and $\hat{k}=k$.
This follows from the bounds on eigenvalues of Hessian matrix and the estimated gradient.
Moreover, we cannot place two estimated modes near any true mode $M_i$
since there is always a saddle point between two (estimated) modes and
this cannot happen due to assumption (M) on the first eigenvalue.
After rearranging the indices, each $\hat{M}_i$ is close to $M_i$ for
all $i=1,\cdots,k$.  Note that $\nabla\hat{f}_n(\hat{M}_i)=0$ so that
Taylor's theorem
implies $$\nabla\hat{f}_n(M_i)=\nabla\hat{f}_n(M_i)-\nabla\hat{f}_n(\hat{M}_i)
= \nabla\nabla \hat{f}_n(M^*_i) (M_i-\hat{M}_i), $$ where $M^*_i$ is a
point between $\hat{M}_i,M_i$.  Thus, $M_i-\hat{M}_i =
\left(\nabla\nabla \hat{f}_n(M^*_i)\right)^{-1} \nabla\hat{f}_n(M_i)$.
By similar technique for proving rates of convergence of the kernel
density estimator, we have $\nabla\hat{f}_n(M_i)
=O(h^2)+O_P\left(\sqrt{\frac{1}{nh^{d+2}}}\right) $. And
$\left\|\left(\nabla\nabla \hat{f}_n(M^*_i)\right)^{-1}\right\|_{\max}$ is
uniformly bounded when $\norm{\hat{f}_n-f}^*_{\infty,2}$ is small. We
conclude $\norm{\hat{M}_i-M_i}_2 =
O(h^2)+O_P\left(\sqrt{\frac{1}{nh^{d+2}}}\right).$ $\Box$

The above theorem shows that every local mode of the intensity
function $f(x)$ will be estimated by the weighted mean
shift algorithm as long as $\norm{\hat{f}_n-f}^*_{\infty,2}$ is sufficiently small. 
By Theorem~\ref{thm::unif}, $\norm{\hat{f}_n-f}^*_{\infty,2}$ is converging
to $0$ so that this result eventually holds.
The rate comes from the bias-variance
decomposition. Note that the variance is
$O_P\left(\sqrt{\frac{1}{nh^{d+2}}}\right)$ since the modes are
defined via the gradient and this is the variance in
estimating gradients.

\begin{thm}
\label{thm::ridges}
Let $\cR,\hat{\cR}_n$ be the ridges of $f$ and $\hat{f}_n$ respectively. Assume (A1-2), (K1-2) and the following two conditions:
\begin{itemize}
\item[(R1)] There exists $\beta,\delta>0$ and $b_1<b_2$ such that $b_1<0$ and $\beta= b_2-b_1$ and for all $x\in\cR\oplus \delta$,
\begin{align*}
\lambda_1(x)>b_2, \quad \lambda_2(x)<b_1,\quad
\norm{\nabla f(x)}_2 \max_{|\alpha|=3}|f^{(\alpha)}(x)|<\frac{\beta^2}{2\sqrt{d}}.
\end{align*}
\item[(R2)] There exists $G_0,G_1>0$ such that 
$$
\{x: \norm{V(x)V(x)^T \nabla f(x)}_{\max}\leq G_0, \lambda_2(x)<G_1\}\subset \cR\oplus \delta,
$$
where $\delta$ is defined in (R1).
\end{itemize}
When $\norm{\hat{f}_n-f}^*_{\infty,3}$ is sufficiently small,
\begin{align*}
d_H\left(\hat{\cR}_n, \cR\right) = O\left(h^2\right)+O_P\left(\sqrt{\frac{\log n}{nh^{d+4}}}\right).
\end{align*}
\end{thm}

{\sc Proof Outline.}  The proof is essentially the same as in
\cite{Genovese2012a}.  In particular, our condition (A1-2) together
with (R1) implies the conditions in \cite{Genovese2012a}.  
We assumed an additional condition (R2) so that, when 
$\norm{\hat{f}_n-f}^*_{\infty,2}$ is sufficiently small, 
the ridge estimator $\hat{\cR}_n\subset \cR \oplus \delta$.
This is because $\hat{f}_n$ and its first,
second derivatives will be bounded near $f$ with high probability and
thus the eigenvalue and projected gradient $V(x)V(x)^T\nabla f(x)$ for
the estimator will be similar to the truth.  Hence, by Theorem 6 and 7
in~\cite{Genovese2012a}, the result follows. $\Box$

\section{Applications}

\begin{figure}
\centering
	\subfigure[Original data]
	{
		\includegraphics[width=1.8 in, height=1.8 in]{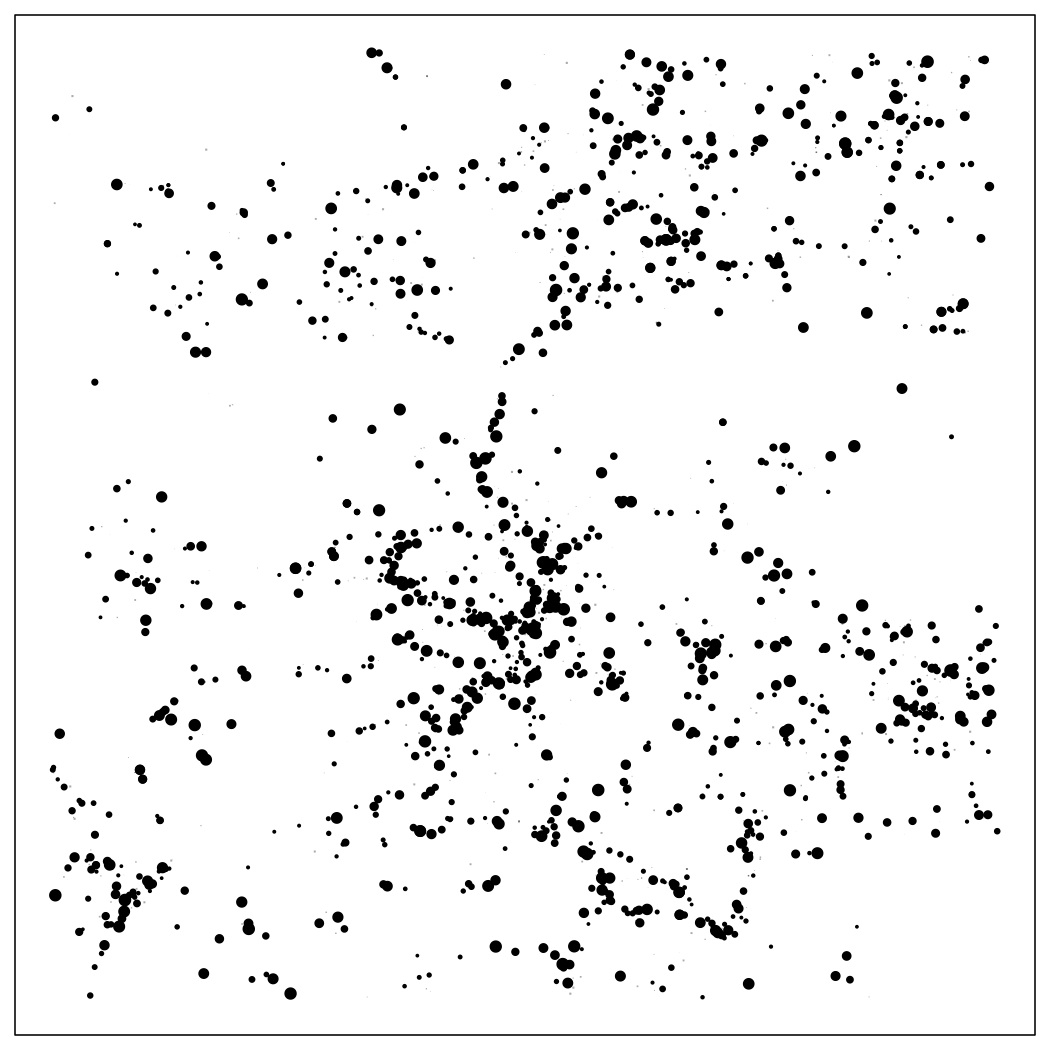}
	}
	\subfigure[Modes and ridges]
	{
		\includegraphics[width=1.8 in, height=1.8 in]{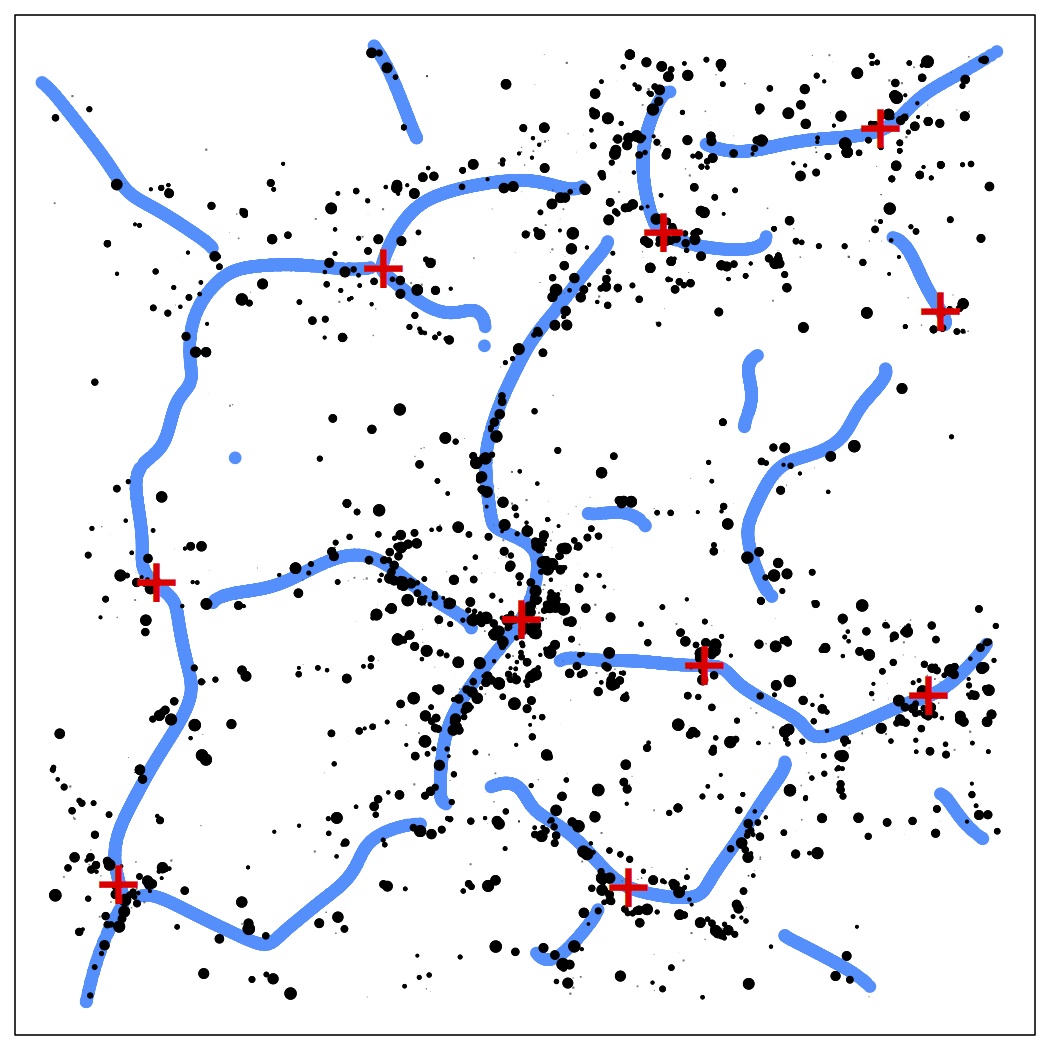}
	}
\caption{The SDSS data. Each black dot is a galaxy. 
The size of the dot is in proportional to the mass of the galaxy.}
\label{fig::ex1}
\end{figure}

We apply our method to two astronomy problems.  The detailed
description of the data can be found in the
supplementary material.

The first problem analyzes data from the Sloan Digit Sky Survey (SDSS),
where each data point is a galaxy at a specific location.
The original data set is three-dimensional, but to ease visualization,
we take a thin slice of the data  to transform it into a two-dimensional problem.
For each galaxy, we also have information on the galaxy's
total luminosity, which can be transformed into a proxy for the galaxy's mass.
We weight each galaxy by the mass and estimate the mass-density field
and find the modes (red crosses) and ridges (blue curves).
In cosmology, the modes correspond to galaxy clusters
and the ridges correspond to filaments.
Figure \ref{fig::ex1} shows the results of our analysis.

The second problem is to identify galaxies in a single image
containing several galaxies.
The data set is a noisy image of four overlapping galaxies. 
We first transform the image into gray scale and convolve with a Gaussian kernel 
with bandwidth $h=4$ to reconstruct the GDF.
We focus only the region with intensity larger than $0.15$.
We apply the weighted mean shift algorithm and perform clustering
to segment the data and construct the connectivity measure for clusters.

\begin{figure}
\centering
	\subfigure[Original image]
	{
		\includegraphics[width=1.85 in, height=1.85 in]{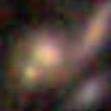}
	}
	\subfigure[Modes and Clustering]
	{
		\includegraphics[width=1.9 in, height=1.9 in]{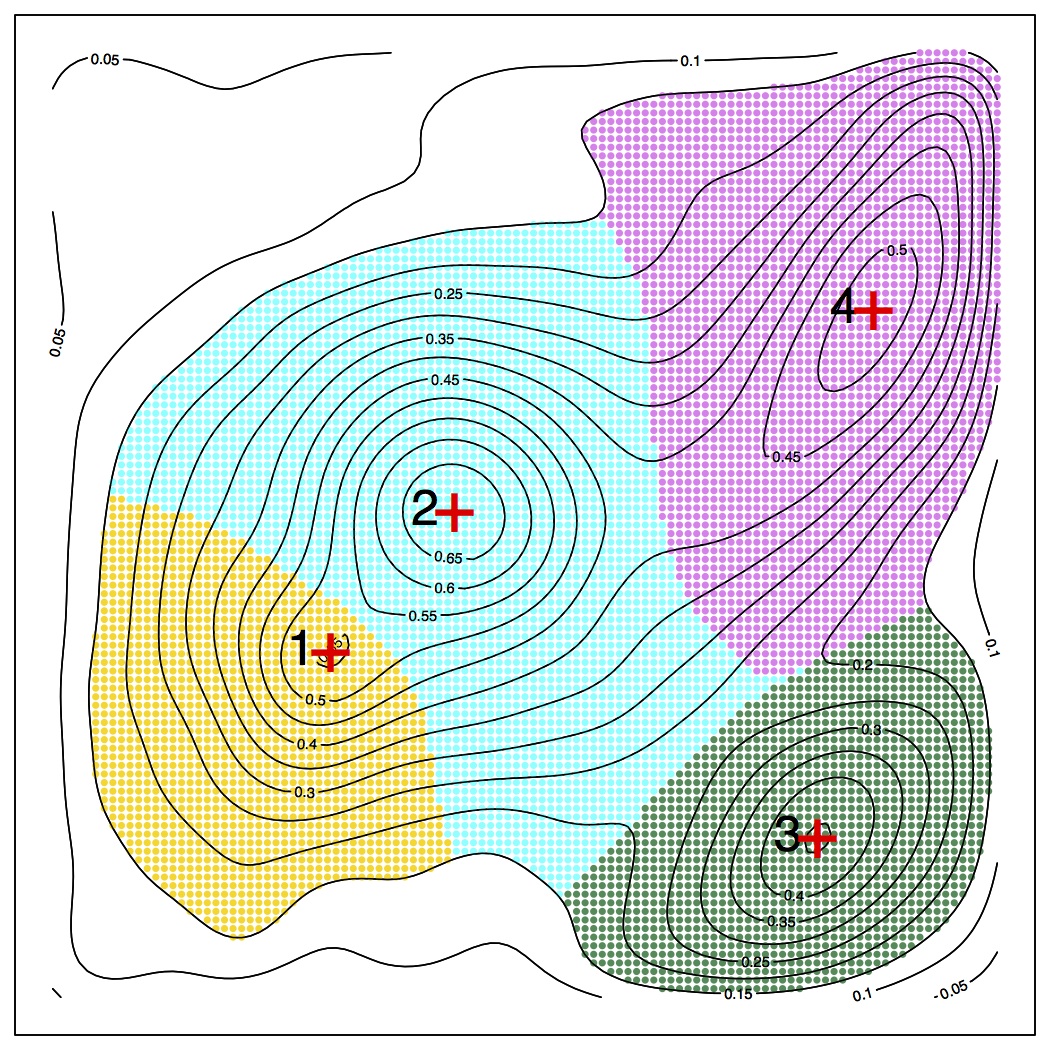} 
	}
	\subfigure[Matrix of connectivity measure]
	{ \centering
\begin{tabular}{rrrrr}
  \hline
 & 1 & 2 & 3 & 4 \\ 
  \hline
1 & -- & 0.25 & 0.10 & 0.10 \\ 
  2 & 0.25 & -- & 0.13 & 0.21 \\ 
  3 & 0.10 & 0.13 & -- & 0.14 \\ 
  4 & 0.10 & 0.21 & 0.14 & -- \\ 
   \hline
\end{tabular}

	}
\caption{The galaxy merger.}
\label{fig::ex2}
\end{figure}

Figure~\ref{fig::ex2} shows the results.
We successfully detect the four galaxies and the weighted mean shift
clustering classifies regions belonging to each galaxy.
The connectivity matrix seems to capture the level of ``interaction'' among
the four galaxies.
For instance, clusters $1$ and $2$ have high connectivity, 
reflecting their large overlap.
In contrast, clusters $2$ and $4$ have only moderate connectivity.

\section{Conclusion}

In this paper, we generalized the mode and ridge estimation 
from densities to generalized densities that
can account for weighting or ``marks.''
We have established convergence rate for estimating modes and ridges
in this case.
Future work will be focused on constructing
confidence sets for the estimated modes and ridges.

\newpage

\appendix

\section{Proof for Theorem 3}

{\sc Proof.}
The proof consists of two steps.
At the first step, we prove that the number of $\cM,\hat{\cM}$ are the same
and each element of $\cM$ correspond to a close element in $\hat{\cM}$.
The second step is to prove the rate of convergence.

We assume
\begin{equation}
\norm{\hat{f}_n-f}^*_{\infty,2}\leq \min\left\{\frac{\lambda_1}{2d}, \frac{\lambda_0}{2}\right\}.
\label{eq::M1}
\end{equation}
(this is what we mean $\norm{\hat{f}_n-f}^*_{\infty,2}$ is sufficiently small)
Note that by Weyl's theorem (Theorem 4.3.1 in \cite{Horn2013}),
\eqref{eq::M1} implies that the eigenvalue difference between 
$\nabla\nabla f(x)$  and its estimator $\nabla\nabla \hat{f}_n(x)$
is bounded by $\frac{1}{2}\lambda_1$. Thus,
eigenvalues of $\nabla\nabla \hat{f}_n(x)$ within $M\oplus \delta$ 
is upper bounded by $\frac{-1}{2\lambda_1}$.
We will use this fact later.

{\bf Step 1.}
Let $\hat{M}_1,\cdots,\hat{M}_{\hat{k}}$ be the estimated local modes. 
Note that the number of estimated modes might be different from the number of true modes.
However, by assumption (M) and \eqref{eq::M1},
we have $\hat{\cM}_n\subset \cM\oplus\delta$.
This can be proved by contradiction since any local mode of $\hat{f}_n$
must have negative first eigenvalue and zero gradient.
Equation \eqref{eq::M1} and assumption (M) force these points 
to be within $\cM\oplus \delta$.

Moreover, we have $\hat{k}=k$.
Note that we cannot place two estimated modes near any true mode $M_i$ 
since there is always a saddle point between two (estimated) modes and this cannot happen
due to the assumption (M) on the first eigenvalue (saddle points have positive first eigenvalue).

{\bf Step 2.}
This part of proof is similar to \cite{Romano1988};
however, our problem is simpler than theirs since we only 
need to find the rate of convergence while they
prove the limiting distributions.
By similar technique, we can prove the limiting distribution as well.

After rearranging the indices, 
each $\hat{M}_i$ is close to $M_i$ for all $i=1,\cdots,k$.
Note that $\nabla\hat{f}_n(\hat{M}_i)=0$ so that Taylor's theorem implies 
$$
\nabla\hat{f}_n(M_i)=\nabla\hat{f}_n(M_i)-\nabla\hat{f}_n(\hat{M}_i) = \nabla\nabla \hat{f}_n(M^*_i) (M_i-\hat{M}_i), 
$$
where $M^*_i$ is a point between $\hat{M}_i,M_i$.
Thus, 
$$
M_i-\hat{M}_i = \left(\nabla\nabla \hat{f}_n(M^*_i)\right)^{-1} \nabla\hat{f}_n(M_i).
$$
The next step is to prove 
\begin{align*}
\mathbb{E}\left(\norm{M_i-\hat{M}_i}_2\right) &= O(h^2)
&\Var\left(\norm{M_i-\hat{M}_i}_2\right) & = O\left(\frac{1}{nh^{d+2}}\right).
\end{align*}

We bound $M_i-\hat{M}_i$ by the following:
\begin{align*}
M_i-\hat{M}_i &=  \left(\nabla\nabla \hat{f}_n(M^*_i)\right)^{-1} \nabla\hat{f}_n(M_i)\\
& \leq \left\|\left(\nabla\nabla \hat{f}_n(M^*_i)\right)^{-1}\right\|_2 \left\|\nabla\hat{f}_n(M_i)\right\|_2.
\end{align*}

We first bound $\left\|\left(\nabla\nabla \hat{f}_n(M^*_i)\right)^{-1}\right\|_2$. 
Note that the $\cL_2$ matrix norm is the largest absoluted eigenvalue; thus,
all we need to do is to bound the eigenvalues of $\left(\nabla\nabla \hat{f}_n(M^*_i)\right)^{-1}$.
Since $M^*_i$ is a point between $M_i,\hat{M}_i$, $M^*_i\in\cM\oplus \delta$. 
Consequently, all eigenvalues of $\nabla\nabla \hat{f}_n(M^*_i)$ must be less or equal to $\frac{-1}{2\lambda_1}$ by assumption (M) and \eqref{eq::M1}.
Therefore, the eigenvalues of $\left(\nabla\nabla \hat{f}_n(M^*_i)\right)^{-1}$ must be bounded by
$2\lambda_1$. This gives the bound on the $\cL_2$ norm.

Now we find the rate of $\left\|\nabla\hat{f}_n(M_i)\right\|_2$.
Since $\nabla f(M_i)=0$,
$$
\mathbb{E}\left(\nabla\hat{f}_n(M_i)\right) = \mathbb{E}\left(\nabla\hat{f}_n(M_i)\right)-\nabla f(M_i)
$$
is the bias of $\hat{f}_n(M_i)$.
By assumptions (A1-2) and (K1), 
the bias is at rate $O(h^2)$  by the same way for find the rate of bias of the kernel density estimator.
Similarly, the covariance matrix is at rate $O\left(\frac{1}{nh^{d+2}}\right)$.

Hence, by multidimensional Chebeshev's inequality,
$$
\nabla\hat{f}_n(M_i) - \mathbb{E}\left(\nabla\hat{f}_n(M_i)\right) = O_P\left(\sqrt{\frac{1}{nh^{d+2}}}\right)
$$
which implies $\nabla\hat{f}_n(M_i) = O(h^2)+O\left(\sqrt{\frac{1}{nh^{d+2}}}\right)$.
Putting altogether, $\norm{M_i-\hat{M}_i}_2=O(h^2)+O_P\left(\sqrt{\frac{1}{nh^{d+2}}}\right)$.
By repeating step 2 for each mode, we conclude the result.
$\Box$

\section{Proof for Theorem 4}
%
%

{\sc Proof.}  The proof is essentially the same as in proving 
Theorem 6 in \cite{Genovese2012a}.  
In particular, our condition (A1-2) together
with (R1) implies all the conditions in \cite{Genovese2012a}.  

Note that they prove that the set $\tilde{\cR} \equiv\hat{\cR}_n\cap  \cR \oplus \delta$
is close to $\cR$ at rate
$$
d_H(\tilde{\cR}, \cR) = O\left(\norm{\hat{f}_n-f}^*_{\infty,2}\right)
$$
as $\norm{\hat{f}_n-f}^*_{\infty,3}$ is sufficiently small.

Here we prove that our additional assumption (R2) implies that 
$\tilde{\cR}=\hat{\cR}$. 
Let $x\in \hat{\cR}$. 
Then we have $\hat{V}_n(x)\hat{V}_n(x)^T \nabla \hat{f}_n(x)=0$.
Thus,
\begin{align*}
\norm{V(x)&V(x)^T \nabla f(x)}_{\max}\\
&=\norm{V(x)V(x)^T \nabla f(x)-\hat{V}_n(x)\hat{V}_n(x)^T \nabla \hat{f}_n(x)}_{\max}\\
&\leq \norm{\left(V(x)V(x)^T-\hat{V}_n(x)\hat{V}_n(x)^T \right) \nabla f(x)}_{\max}+
\norm{\hat{V}_n(x)\hat{V}_n(x)^T\left(\nabla f(x)-\nabla \hat{f}_n(x)\right)}_{\max}\\
&\leq \norm{V(x)V(x)^T-\hat{V}_n(x)\hat{V}_n(x)^T }_{\max}C_1+\norm{\nabla f(x)-\nabla \hat{f}_n(x)}_{\max}C_2
\end{align*}
where $C_1,C_2$ are two constants that is independent of $x$. 
The existence of $C_1,C_2$ comes from the fact that 
$\nabla f(x)$ and $\hat{V}_n(x)\hat{V}_n(x)^T$
are uniformly bounded for all $x \in \cR$.
Note that since $\hat{V}_n\hat{V}_n^T$ is a projection matrix,
its max norm is uniformly bounded by $1$.
The first term $\norm{V(x)V(x)^T-\hat{V}_n(x)\hat{V}_n(x)^T }_{\max}$
can be bounded by Davis-Kahan's theorem~\cite{Luxburg2007} 
at rate $\norm{\hat{f}_n-f}^*_{\infty,2}$ and the second term
is at rate $\norm{\hat{f}_n-f}^*_{\infty,1}$.
Accordingly, as $\norm{\hat{f}_n-f}^*_{\infty,3}$ is small,
any point $x\in \hat{\cR}$ satisfies 
$\norm{V(x)V(x)^T \nabla f(x)}_{\max}\leq G_0, \lambda_2(x)<G_1$. 
Thus, $\hat{\cR}\subset \cR\oplus\delta$ so that
$\tilde{\cR}=\hat{\cR}$. 

Now by Theorem 6 in \cite{Genovese2012a},
$$
d_H(\hat{\cR}, \cR) = O\left(\norm{\hat{f}_n-f}^*_{\infty,2}\right)
$$
and by Theorem \ref{thm::unif}, we conclude the result.

 $\Box$

\section{Preprocessing and Description of The Data}

\subsection{The SDSS Data}
The data in the Sloan Digit Sky Survey (SDSS) can be found in \url{http://www.sdss3.org/}.
We use the Main Sample Galaxies (MSGs) spectroscopic data in the data release 9.
Each observation is a galaxy conatining the following four features:
\begin{itemize}
\item[1.] RA    = right ascension (i.e., longitude, in degrees [0 to 360])
\item[2.] DEC   = declination (i.e., latitude, in degrees [-90 to 90])
\item[3.] z     = spectroscopic redshift (precisely estimated proxies for distance)
\item[4.] r     = Petrosian magnitude, r band (for completeness only)
\end{itemize}
The first three features (RA, DEC, z) relates to the spatial position of the galaxy
and the Petrosian magnitude is a luminosity measure associated with the mass.

We select a thin slice of the universe to analyze:
\begin{align*}
RA&\in [155,185] &DEC&\in [35,65] & z &\in [0.110,0.115].
\end{align*}
The mass of a galaxy is given by the following formula (\url{http://www.sdss.org/DR7/tutorials/getdata/index.html}):
$$
{\sf MASS} = r - 5\log (4.28\times 10^{8}\times z),
$$
where $r$ is the Petrosian magnitude and $z$ is the redshift.

We use the {\sf MASS} as $Y$ and the spatial location $X=(RA,DEC)$
to conduct our analysis. Note that since we pick a thin slice, 
we can neglect the dimension of $z$.

\subsection{The Galaxy Merger}
The galaxy merger image is taken from the galaxy zoo (\url{http://quenchtalk.galaxyzoo.org/#/subjects/AGS000007y})
with label `AGS000007y'.

For the image data, the statistical model is given by
$$
Y_i = g(X_i)+\epsilon_i,\qquad i=1,\cdots,n,
$$
where $\epsilon_i$'s are IID mean $0$ noises and $X_i$ is from an uniform grid.
Since $X_i$ is on an uniform grid, the density function is constant,
the function $g$ is the same as the GDF.
Then the estimator 
$$
\hat{g}_n(x) = \frac{1}{nh^d}\sum_{i=1}^n Y_iK\left(\frac{x-X_i}{h}\right)
$$
is an estimate to $g(x)$. And all the analysis for GDF can be applied
to the image data.

\bibliographystyle{abbrvnat}
\bibliography{wSCMS.bib}

---
\end{document}